\documentclass[twocolumn,aps,prl,showpacs,superscriptaddress]{revtex4}
\usepackage{graphicx}
\newcommand{\YBCO}{YBa$_2$Cu$_3$O$_{6+x}$}
\newcommand{\Tc}{$T_c$}
\newcommand{\um}{$\mu$m}
\newcommand{\Flat}{F_{\rm lat}}
\newcommand{\Fpin}{F_{\rm pin}}
\newcommand{\Fig}[1]{Fig.~\ref{#1}}
\newcommand{\Figs}[1]{Figs.~\ref{#1}}

\begin{document}

\author{Lan Luan}
\author{Ophir M. Auslaender}
\affiliation{Geballe Laboratory for Advanced Materials, Stanford University, Stanford, CA 94305, USA}
\author{Douglas A. Bonn}
\author{Ruixing Liang}
\author{Walter N. Hardy}
\affiliation{Department of Physics and Astronomy, University of British Columbia, Vancouver, BC, Canada V6T 1Z1}
\author{Kathryn A. Moler}
\email[Corresponding author: ]{kmoler@stanford.edu}
\affiliation{Geballe Laboratory for Advanced Materials, Stanford University, Stanford, CA
94305, USA}

\title{Magnetic-force-microscope study of interlayer ``kinks'' in individual vortices in underdoped cuprate \YBCO\ superconductor}

\begin{abstract}
We use magnetic force microscopy to both image and manipulate individual vortex lines threading single crystalline YBa$_2$Cu$_3$O$_{6.4}$, a layered superconductor. We find that when we pull the top of a pinned vortex, it may not tilt smoothly. Sometimes, we observe a vortex to break into discrete segments that can be described as short stacks of pancake vortices, similar to the ``kinked" structure proposed by Benkraouda and Clem. Quantitative analysis gives an estimate of the pinning force and the coupling between the stacks. Our measurements highlight the discrete nature of stacks of pancake vortices in layered superconductors.
\end{abstract}


\maketitle

Magnetic field penetrates superconductors in the form of vortices, each carrying one magnetic flux quantum, $\Phi_0\equiv h/2e$. In the highly anisotropic cuprates, where the c-axis penetration depth ($\lambda_c$) is much larger than the in-plane penetration depth ($\lambda_{ab}$), the three-dimensional vortex can be treated as a stack of two-dimensional, magnetically coupled, ``pancake'' vortices \cite{Clem1991,Artemenko_1990}, with weak interlayer Josephson coupling \cite{Clem04,bulaevskii1992}. Rich physics arises from the competition between thermal energy, vortex-vortex interactions, pinning and the interlayer coupling. While numerous studies have been done on vortex-matter thermodynamics, mostly by measuring macroscopic properties \cite{Blatter1994, Beidenkopf07, Figueras06},  work on individual vortices is scarce. Magnetic force microscopy (MFM) allows us to manipulate individual vortices\cite{moser_1998} with a high level of control \cite{straver_2008,Auslaender08}. Here we use MFM to directly probe the pinning energy and the interlayer coupling energy, which determine the shape of a vortex and the nature of its motion.
\begin{figure}[b]
\includegraphics[width=3.3in]{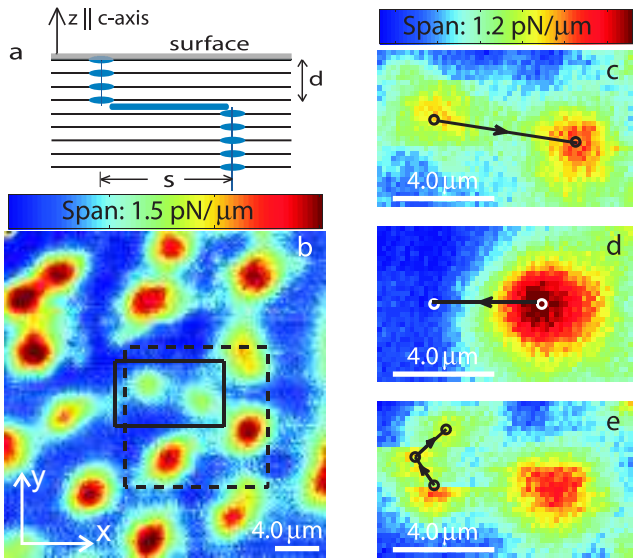}
\caption{\label{fig_annicreat}MFM images showing the annihilation and creation of kinked stacks of pancake vortices, which appear as pairs of sub-$\Phi_0$, isolated features. %
{\bf(a)} Cartoon of a side view of a kinked pancake stack (ellipses), including the axis of the vortex (vertical lines) and the interlayer Josephson vortex (thick blue line). $d$ is the depth of the kinked structure and $s$ is the lateral separation between the stacks. Also depicted are the Cu-O layers (horizontal lines). Not drawn to scale. %
{\bf(b)} Initial configuration of vortices after field cooling through \Tc\ to $T=5.4$~K. Scan height $z = 1.05~\mu$m. The solid frame shows the scan area for \Figs{fig_annicreat}(c)-(e) and highlights a pair of separated partial stacks. The dashed frame shows the scan area for \Fig{fig_fit}. %
{\bf(c-e)} Scans at $T=12$~K of the two pancake stacks in the solid frame in (b). The arrows show the tip path used for manipulating the stacks as described in the text. (c) Scan before annihilation ($z = 0.93~\mu$m), (d) scan after annihilation and before creation ($z=1.24~\mu$m) and (e) scan after creation ($z=1.24~\mu$m). In (e) the tip starts scanning from the bottom left corner and is incremented along $+\hat y$ after each raster period. The vortex stack on the left jumps as the tip scans over it. The dots and the arrows show the positions where the stack was trapped temporarily and the trajectory of its motion. Here, the force exerted from the tip is much smaller than required to move a regular vortex, indicating the stack configuration is unstable.}
\end{figure}

In a layered superconductor a vortex subject to shear can theoretically break into separate straight stacks of pancakes, to create a kinked structure , instead of tilting \cite{Benkraouda1996} [\Fig{fig_annicreat}(a)]. Previous researchers have identified such kinked vortices by direct imaging \cite{Guikema08,Beleggia04,Grigorenko2002}, proposed a model for kinked stacks in the presence of pinning \cite{Benkraouda1996,Guikema08}, and discussed the interaction between pancake vortices and Josephson vortices \cite{Grigorenko2001,vlasko-vlasov02}. Here we use MFM to directly test the picture of kinked vortices interacting with pinning and to measure the coupling between the stacks composing a single vortex in an underdoped \YBCO\ (YBCO) single crystal. We sometimes observe pairs of well separated magnetic features on the sample surface carrying sub-$\Phi_0$ flux. Using the magnetic tip of the MFM for manipulation, we combine pairs of features to create regular, $\Phi_0$, vortices, verifying the kinked stack picture. As further corroboration, we split regular $\Phi_0$-vortices by pulling them apart to create kinks. We measure the required force, obtaining an estimate for the attractive interaction between stacks. The result agrees well with a model of magnetic coupling augmented by the line-tension of the Josephson string connecting the stacks.

The YBCO single crystal was grown by the self-flux method in BaZrO$_3$ crucibles \cite{Liang1998}, mechanically detwinned and annealed. The platelet-shaped crystal (face parallel to the ab-plane, dimensions $0.7~{\rm mm}\times0.7~{\rm mm}$) is 100~\um~ thick. The anisotropy of YBCO increases as the superconducting transition temperature \Tc\ drops with reduced doping. For our sample, $T_c=21$~K (transition width $2$~K), implying $x\approx0.4$ and therefore $\lambda_{ab}\approx0.36$~\um\ and $\gamma \equiv \lambda_c/\lambda_{ab} \approx 75$ at zero temperature \cite{Hosseini04,Broun07,Liang05}, comparable to that of the highly anisotropic superconductor Bi$_2$Sr$_2$CaCu$_2$O$_8$ ($\gamma \approx 60-250$)\cite{Blatter1994}.

Our measurements are performed in a variable-temperature MFM in frequency modulation mode \cite{Albrecht_1991}. A magnetic tip at the end of a cantilever \cite{Cantilever} is scanned at a constant height $z$ above the sample surface which is experimentally determined by obtaining a parabolic fit to dozens of touching down positions. The force between the tip and the sample induces a shift of the resonant frequency of the cantilever, $f_0$, which we measure. Subtracting a $z$-dependent offset, we obtain the contribution of the tip-vortex interaction, $\Delta f$, which gives information on the tip-vortex force, $\partial F_z / \partial z=-2k \Delta f/f_0$, where $k=2.7 \pm 0.1$~N/m is the cantilever spring constant \cite{SpringConstant} and $f_0=59.040$~kHz. In our scans, the tip moves back and forth along the x-axis, then, after completing one period of motion (duration of a few seconds), it is incremented along either $+\hat y$ or $-\hat y$. The choice $\pm$ differs from scan to scan.

The force exerted by the tip on the sample is generally regarded as a drawback of MFM. Here, we magnetize the tip by placing near it a permanent magnet with the polarity chosen to give attractive tip-vortex force. We deliberately use the lateral components of this force, $\vec\Flat$, to overcome the pinning force, $\Fpin$, in order to manipulate individual vortices. We tune $\Flat\equiv\left|\vec\Flat\right|$ by varying the scan height $z$. We first image at a height where the tip force is insufficient to perturb the vortices, which are held static by $\Fpin$, i.e. $\Flat\left(z\right)< \Fpin\left(T\right)$. For manipulation, we reduce $z$ to increase $\Flat$ [in the model described below, $\max{\left(\Flat\right)} \equiv\Flat^{\max} \sim z^{-2}$ for $z\gg\lambda_{ab}$]. When $\Flat^{\max}\left(z\right)>\Fpin\left(T\right)$ we can manipulate a vortex. Temperature, which reduces $\Fpin$, gives extra control \cite{Auslaender08}.

For low vortex density, we cooled the sample in a magnetic field of $0.5\cdot10^{-4}$~T along the crystal's c-axis. \Fig{fig_annicreat}(b) shows an image acquired at $T=5.4$~K, in which vortices appear as peaks. Most vortices give the same peak height, as expected, since they each should carry a flux of exactly $\Phi_0$. However, some peaks have weaker amplitude and always appear in pairs [e.g. in the solid framed region in \Fig{fig_annicreat}(b)], indicating the flux associated with each member is less than $\Phi_0$. Previous work \cite{Guikema08} suggests that these peaks originate from kinked stacks of pancake vortices forming one $\Phi_0$-vortex. To test this model, we annihilated kinks and then recreated them [\Figs{fig_annicreat}(c)-(e)]. For the manipulation we heated the sample to $T=12$~K, to reduce $\Fpin$. Then, after locating two distinct partial stacks [\Fig{fig_annicreat}(c)], we tried to pull one towards the other by moving the tip along the line between them. We repeated this several times, reducing $z$ for each new attempt. While driving back to the starting position for a new attempt, we made sure to retract the tip, reducing $\Flat^{\max}$ and with it the chance of accidentally perturbing the vortex. We found that after we succeeded to move one vortex stack, it combined with its partner to form a $\Phi_0$-vortex with good rotational symmetry, suggesting the stacks are well aligned [\Fig{fig_annicreat}(d)]. As an additional test, we pulled the vortex apart again by moving the tip away from the center of the combined vortex at $\vec{R_i}$ [\Fig{fig_annicreat}(d)] and successfully created two distinct stacks using the same $z$ and $T$ as for the annihilation [\Fig{fig_annicreat}(e)]. The newly created partial stack was not always stable, as indicated by the vortex jumps in \Fig{fig_annicreat}(e).
\begin{figure}[pb]
\includegraphics[width=3.3in]{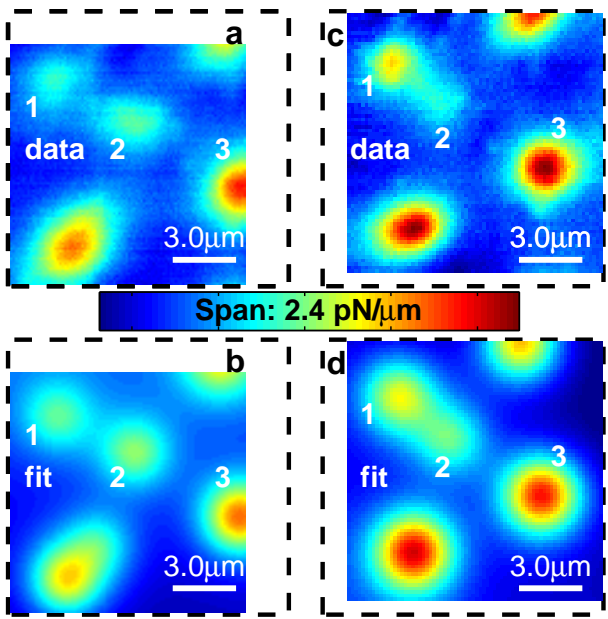}
\caption{\label{fig_fit} Scans (a,c) and fits (b,d) of vortices before and after manipulation. The dashed frames mark the same area as the dashed frame in \Fig{fig_annicreat}(a). %
{\bf(a)} Scan at $T=5.3$~K, $z=0.94$~\um, of a field cooled vortex configuration. %
{\bf(b)} Fit to scan in (a) using the monopole-monopole model. The fitted amplitudes are: $A_1=0.69\pm 0.01~{\rm pN}/\mu{\rm m}$, $A_2=0.86\pm 0.01~{\rm pN}/\mu{\rm m}$, $A_3=1.55\pm 0.01~{\rm pN}/\mu{\rm m}$ (errors denote 95\% confidence intervals), giving $A_1 + A_2\approx A_3$. For this pair the fit gives $d=0.6\lambda_{ab}$ and $s=4.1$~\um. %
{\bf(c)} Scan of vortices at $T=5.3$~K, $z=0.84$~\um, after combining and re-separating feature 1 and 2. Both the relative signal strength and the separation between the two stacks changed because of the manipulation. %
{\bf(d)} Fit to scan in (c). Fitted amplitudes: $A_1=1.28\pm 0.03~{\rm pN}/\mu{\rm m}$, $A_2=0.94\pm 0.03~{\rm pN}/\mu{\rm m}$, $A_3=2.00\pm 0.02~{\rm pN}/\mu{\rm m}$. As in (b), $A_1 + A_2\approx A_3$. For this pair: $d=0.9\lambda_{ab}$, $s=3.2$~\um.}
\end{figure}

To confirm that the two stacks we manipulated compose one vortex, we fit the scan before annihilation [\Fig{fig_fit} (a),(b)]. The model we used is based on the fact that, for $z \gg\lambda_{ab}$, the magnetic field from a stack of pancake vortices is approximately the field from a magnetic monopole residing $\lambda_{ab}$ beneath the superconducting surface \cite{clem_1994}. The flux associated with the vortex stack is $\beta \Phi_0$. For a stack extending from depth $d$ up to the surface, $\beta= 1- e^{-d/\lambda_{ab}}$. For a semi-infinite stack extending from depth $d$ down: $\beta = e^{-d/\lambda_{ab}}$ \cite{Guikema08}. We find that our tip can be well approximated by a long narrow cylinder magnetized along its axis, $\hat{z}$. The MFM signal from a collection of vortex stacks is then given by the ``monopole-monopole" model:
$\partial F_z / \partial z = \sum_i A_i \frac{1-\frac12\left(\vec{R}-\vec{R}_i\right) ^2/\left(z+h_0\right)^2} {\left[1+\left(\vec{R}-\vec{R_i}\right) ^2 / \left(z+h_0\right) ^2 \right] ^{5/2}}$,
where $i$ enumerates the distinct vortex features in a scan, $\vec{R}$ is the in-plane position of the tip; $A_i = \beta \tilde{m} \Phi_0\left(z+h_0\right)^{-3}/\pi$ and $h_0 = \lambda_{ab} + d_{\rm offset}$ where $\tilde{m}$ is the dipole moment per unit length, $d_{\rm offset}$ is the offset due to the tip geometry and the non-superconducting layer on the surface of the superconductor \cite{Auslaender08}). The fit in \Fig{fig_fit}(b) gives $A_1 + A_2 \approx A_3 $, confirming that two adjacent partial vortex stacks add up to one regular vortex.

Other experimental observations provide further insight about the kinked stacks. When we recombined and re-separated the same pair of stacks repeatedly, we could only manipulate one member. Presumably, this stack was the finite top stack. Furthermore, both the separation between the stacks and the signal amplitudes changed in the annihilation-creation process (e.g. \Fig{fig_fit}), reasserting that pinning is important and that the tip allows the dragged pancakes to explore the pinning environment. Finally, partial pairs were rare [only one in \Fig{fig_annicreat}(b)]. However, most other vortices had irregular shape, which we believe is due to misalignment of the pancakes, too small to be resolved as distinct stacks because of the relatively large $\lambda_{ab}$. This irregularity tended to diminish after dragging, in support of the picture that pinning hinders pancake stacks from aligning.
\begin{figure}[pb]
\includegraphics[width=3.3in]{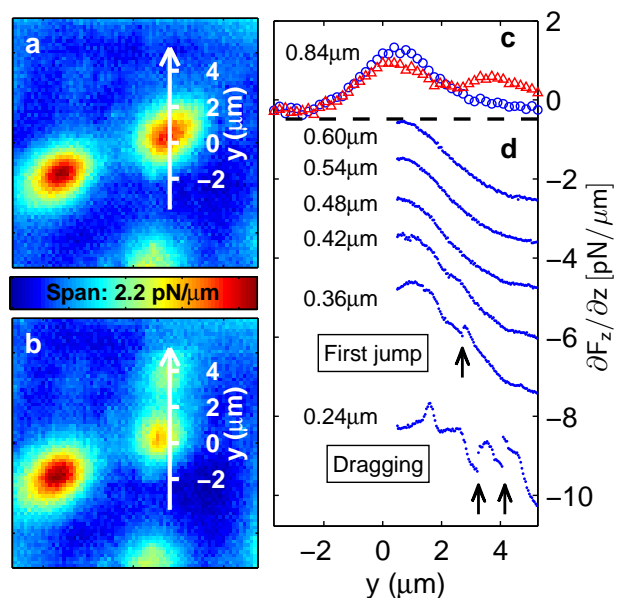}
\caption{\label{fig_kink} Creating kinks in a straight stack. %
{\bf(a)} Image of two regular, field cooled, vortices at $T=5.3$~K ($z=0.80~\mu$m). The white line marks the location of the line-cut plotted in (c) and the path for the line-scans in (d). %
{\bf(b)} Image of a regular vortex [the one on the left in (a)] and the newly created kinked-stack pair, created by the sequence of line scans in (d). The white line is the same as in (a).  Scan at $T=5.3$~K, $z= 0.80~\mu$m. %
{\bf(c)} Circles and triangles show points extracted along the white line in (a) and (b), respectively. %
{\bf(d)} Line scans acquired with the tip moving over the white lines in (a) and (b) starting from $y=0$ ($T=12$~K). $z=0.60,~0.54,~0.48,~0.42$~\um: the vortex was stationary at approximately $y=0$, with the signal dropping as the tip moved away. $z=0.36~\mu$m: the first jump of the vortex towards the tip (tip position when this happens is marked by an arrow, roughly when the lateral force peaks (\Fig{fig_forceCal})), as indicated by the abrupt signal increase due to the increased tip-vortex interaction. $z= 0.24~\mu$m: the vortex was dragged by the tip. The arrows highlight the discontinuities of the trace due to vortex motion. We believe that other sharp features in the trace originate from bumps on the surface which would deflect the tip as we further lowered the scan height.}
\end{figure}

\begin{figure}[pbt]
\includegraphics[width=3.3in]{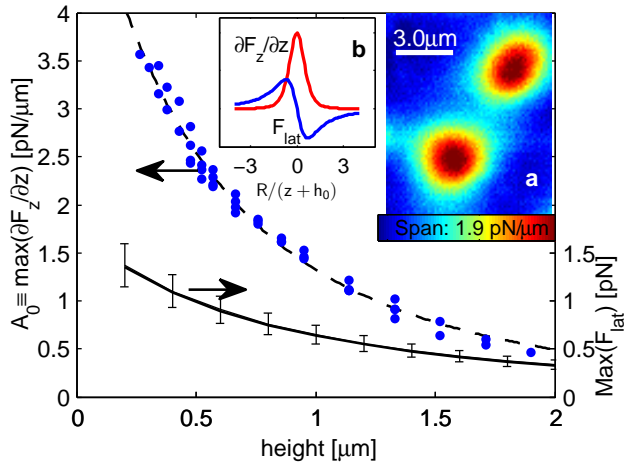}
\caption{\label{fig_forceCal}Calibration of the tip to extract $\Flat^{\max}$. At each $z$ we acquired a scan of two immobile vortices at $T=5.0$~K [insert (a) -- an example for $z=0.85~\mu$m]. We fit each scan to the monopole-monopole model and plot the peak amplitude, $A_0\equiv\max\left(\partial F_z/\partial z\right)$ (main panel, left ordinate). We then fit the result to $\left( \tilde{m} \Phi_0 / \pi \right) / \left(z+h_0\right)^3$, the dependence of the peak height on $z$ in the model (dashed line). The fit yields $\tilde{m}=22\pm 3~{\rm nA}\cdot$m and $h_0=1.55\pm 0.10~\mu$m\cite{doffset}. Within the model $\Flat^{\max}=\left( \tilde{m} \Phi_0 / 3\sqrt{3}\pi \right) / \left(z+h_0\right)^2$ (solid line, right ordinate). Error bars show the 95\% confidence interval from the fit parameter uncertainty.
Note that we performed the manipulation at elevated temperature (e.g. $T=10$~K for \Fig{fig_kink}), which leads to an additional 10\% systematic error in $\Flat^{\max}$, due to the increase of $\lambda_{ab}$ \cite{Hosseini04}. %
Insert (b) shows $\partial F_z / \partial z$ and $\Flat$ in the model as a function of $R$ in units of $\left(z+h_0\right)$.  Note that $\Flat\left(0\right)=0$ and reaches $\Flat^{\max}$ at $R=\left(z + h_0 \right)/\sqrt{2}$, roughly when $\partial F_z/\partial z=\max\left(\partial F_z/\partial z \right)/2$.}
\end{figure}

The deliberate annihilation and recreation of partial vortex pairs confirms the picture of kinked pancake stacks interacting with local pinning. To provide quantitative information on the coupling between the pancakes composing a single vortex, we measure the force required to create a kink (\Fig{fig_kink}). To that end, we moved the tip repeatedly along a line leading away from a regular vortex at constant $z$, approaching closer to the surface for each new line-scan to increase $\Flat$ [\Fig{fig_kink}(d)]. We estimate $\Flat$ from the monopole-monopole model (\Fig{fig_forceCal}). For large $z$, the vortex remained unperturbed, implying $\Flat^{\max}(z) < \Fpin(T)$. We estimate $\Fpin(T)$ from the largest $z$ at which we observe vortex motion, manifesting itself as discontinuities in the scan trace well above the noise level (0.05~pN/$\mu{\rm m}$). For example, at $T=10$~K we observed first motion at $z = 0.36~\mu{\rm m}$, giving $\Fpin(T=10~{\rm K}) \approx\Flat^{\max} \left( z = 0.36~\mu{\rm m}\right) \approx 1.2$~pN. For lower scans (e.g. $z=0.24~\mu$m), the tip dragged part of the vortex to a new position, creating two distinct stacks [\Fig{fig_kink}(b)]. In order to pull a vortex apart, $\Flat$ has to overcome both $\Fpin$ and the restoring force $F_{\rm el}$, which binds the two partial stacks together. At the position where the partial stack stopped following the tip, $\Flat^{\max}\approx\Fpin + F_{\rm el}$. Thus, the measured restoring force can be estimated from: $\Flat^{\max}(z= 0.24~\mu{\rm m})-\Flat^{\max}(z= 0.36~\mu{\rm m})$ giving $F_{\rm el}\approx 0.1$~pN for a stack separation of $s=3.3~\mu$m, as identified by fitting \Fig{fig_kink}(b) to the monopole-monopole model.

The restoring force, $F_{\rm el}$, has two contributions, both attractive: the magnetic coupling between pancakes in different layers and the Josephson-string line-tension. The former is obtained by summing over the magnetic interaction between all the pancakes in the two partial stacks. Benkraouda and Clem (BC) \cite{Benkraouda1996} calculated this force for two stacks of equal length, long on the scale of $\lambda_{ab}$.
In our case, the length of the top stack, $d$, is of order $\lambda_{ab}$. For $s \gg \lambda_{ab}$ we obtain the BC result, suppressed by a factor of approximately $1- e ^{-d/\lambda_{ab}}$, to give:
$F_{\rm mag}(s)=-\partial E_{\rm mag}/\partial s\approx\left(\Phi_0/4\pi\lambda_{ab}\right)^2\left[{\lambda_{ab}}/{s}-e^{-s/\lambda_{ab}}\left(1+{\lambda_{ab}}/{s}\right)\right]\left[1-e^{-d/\lambda_{ab}}\right]$.
The line tension of a Josephson string for $\lambda_{ab} < s < \lambda_c $ is: $F_J=-\partial E_J/\partial s\approx\left(\Phi_0/4\pi\right)^2 \left(\lambda_{ab}\lambda_c\right)^{-1}$ \cite{Clem04}. Given $\lambda_{ab}=0.40~\mu$m and $\lambda_c=31.6~\mu$m at $T=10$~K \cite{Broun07,Hosseini04}, $s=3.3~\mu$m and $d=0.5\lambda_{ab}$ [from fitting \Fig{fig_kink}(b)], we find $F_{\rm mag}=0.09$~pN and $F_J\approx0.02$~pN. Adding the two gives $F_{\rm tot}= F_J + F_{\rm mag}=0.11$~pN, in good agreement with our estimate from the measurement.

We manipulated 20 vortices at various temperatures near $T_c/2$. For some we thermal-cycled above \Tc\ and slightly changed the applied magnetic field for a different initial vortex configuration. We successfully created and observed the kinked structure in two vortices. In the remaining cases we could drag the top of the vortex but did not observe kinking. This is not unexpected: by MFM we can only manipulate pancakes that lie at most a few $\lambda_{ab}$ beneath the surface because of the exponential suppression of the force from the tip. Imaging is also limited, because the MFM's resolution is set by $z+h_0$, the scale on which the field from a stack decays. The low rate of creating observable kinks in the limited volume defined by $d$ and $s$ (10\%) and of observing as field-cooled partial stacks (one pair in 3 thermal-cycles) reasserts that the balance between local pinning and $F_{\rm el}$ is crucial for determining the alignment of the pancakes composing a vortex. It also suggests that the distribution of pinning cites has spacial variation, but is not strongly inhomogeneous.

To conclude, by using MFM for imaging and manipulating individual vortices, we have verified that separated pancake stacks in the presence of pinning are the appropriate description for vortices in highly anisotropic type-II superconductors. We have measured the force required to depin a pancake stack and to create the kinked structure, providing quantitative information on the coupling between pancake stacks in a single vortex. The experiment highlights single vortex mechanical properties, hard to extract from macroscopic measurements. The technique, manipulating individual vortices by MFM, opens unique possibilities to study interacting many-body systems \cite{Nelson1993}, as well as to address open questions in vortex matter, e.g. the cutting barrier for vortices and vortex entanglement \cite{Nelson1988, olson_reichhardt2004}.

Acknowledgment: The authors would like to thank H.~Bluhm and B.~Kalisky for helpful discussions. The work is supported on DOE contract no. DE-AC02-76SF00515.

\newcommand{\noopsort}[1]{} \newcommand{\printfirst}[2]{#1}
  \newcommand{\singleletter}[1]{#1} \newcommand{\switchargs}[2]{#2#1}


\begin{thebibliography}{29}
\expandafter\ifx\csname natexlab\endcsname\relax\def\natexlab#1{#1}\fi
\expandafter\ifx\csname bibnamefont\endcsname\relax
  \def\bibnamefont#1{#1}\fi
\expandafter\ifx\csname bibfnamefont\endcsname\relax
  \def\bibfnamefont#1{#1}\fi
\expandafter\ifx\csname citenamefont\endcsname\relax
  \def\citenamefont#1{#1}\fi
\expandafter\ifx\csname url\endcsname\relax
  \def\url#1{\texttt{#1}}\fi
\expandafter\ifx\csname urlprefix\endcsname\relax\def\urlprefix{URL }\fi
\providecommand{\bibinfo}[2]{#2}
\providecommand{\eprint}[2][]{\url{#2}}

\bibitem[{\citenamefont{Clem}(1991)}]{Clem1991}
\bibinfo{author}{\bibfnamefont{J.~R.} \bibnamefont{Clem}},
  \bibinfo{journal}{Phys. Rev. B} \textbf{\bibinfo{volume}{43}},
  \bibinfo{pages}{7837} (\bibinfo{year}{1991}).

\bibitem[{\citenamefont{Artemenko and Kruglov}(1990)}]{Artemenko_1990}
\bibinfo{author}{\bibfnamefont{S.~N.} \bibnamefont{Artemenko}}
  \bibnamefont{and} \bibinfo{author}{\bibfnamefont{A.~N.}
  \bibnamefont{Kruglov}}, \bibinfo{journal}{Phys. Lett. A}
  \textbf{\bibinfo{volume}{143}}, \bibinfo{pages}{485} (\bibinfo{year}{1990}).

\bibitem[{\citenamefont{Clem}(2004)}]{Clem04}
\bibinfo{author}{\bibfnamefont{J.~R.} \bibnamefont{Clem}}, \bibinfo{journal}{J.
  Supercond.} \textbf{\bibinfo{volume}{17}}, \bibinfo{pages}{613}
  (\bibinfo{year}{2004}).

\bibitem[{\citenamefont{Bulaevskii et~al.}(1992)\citenamefont{Bulaevskii,
  Ledvij, and Kogan}}]{bulaevskii1992}
\bibinfo{author}{\bibfnamefont{L.~N.} \bibnamefont{Bulaevskii}},
  \bibinfo{author}{\bibfnamefont{M.}~\bibnamefont{Ledvij}}, \bibnamefont{and}
  \bibinfo{author}{\bibfnamefont{V.~G.} \bibnamefont{Kogan}},
  \bibinfo{journal}{Phys. Rev. B} \textbf{\bibinfo{volume}{46}},
  \bibinfo{pages}{11807} (\bibinfo{year}{1992}).

\bibitem[{\citenamefont{Blatter et~al.}(1994)\citenamefont{Blatter, Feigel'man,
  Geshkenbein, Larkin, and Vinokur}}]{Blatter1994}
\bibinfo{author}{\bibfnamefont{G.}~\bibnamefont{Blatter}},
  \bibinfo{author}{\bibfnamefont{M.~V.} \bibnamefont{Feigel'man}},
  \bibinfo{author}{\bibfnamefont{V.~B.} \bibnamefont{Geshkenbein}},
  \bibinfo{author}{\bibfnamefont{A.~I.} \bibnamefont{Larkin}},
  \bibnamefont{and} \bibinfo{author}{\bibfnamefont{V.~M.}
  \bibnamefont{Vinokur}}, \bibinfo{journal}{Rev. Mod. Phys.}
  \textbf{\bibinfo{volume}{66}}, \bibinfo{pages}{1125} (\bibinfo{year}{1994}).

\bibitem[{\citenamefont{Beidenkopf et~al.}(2007)\citenamefont{Beidenkopf,
  Verdene, Myasoedov, Shtrikman, Zeldov, Rosenstein, Li, and
  Tamegai}}]{Beidenkopf07}
\bibinfo{author}{\bibfnamefont{H.}~\bibnamefont{Beidenkopf}},
  \bibinfo{author}{\bibfnamefont{T.}~\bibnamefont{Verdene}},
  \bibinfo{author}{\bibfnamefont{Y.}~\bibnamefont{Myasoedov}},
  \bibinfo{author}{\bibfnamefont{H.}~\bibnamefont{Shtrikman}},
  \bibinfo{author}{\bibfnamefont{E.}~\bibnamefont{Zeldov}},
  \bibinfo{author}{\bibfnamefont{B.}~\bibnamefont{Rosenstein}},
  \bibinfo{author}{\bibfnamefont{D.}~\bibnamefont{Li}}, \bibnamefont{and}
  \bibinfo{author}{\bibfnamefont{T.}~\bibnamefont{Tamegai}},
  \bibinfo{journal}{Phys. Rev. Lett.} \textbf{\bibinfo{volume}{98}},
  \bibinfo{pages}{167004} (\bibinfo{year}{2007}).

\bibitem[{\citenamefont{Figueras et~al.}(2006)\citenamefont{Figueras, Puig,
  Obradors, Kwok, Paulius, Crabtree, and Deutscher}}]{Figueras06}
\bibinfo{author}{\bibfnamefont{J.}~\bibnamefont{Figueras}},
  \bibinfo{author}{\bibfnamefont{T.}~\bibnamefont{Puig}},
  \bibinfo{author}{\bibfnamefont{X.}~\bibnamefont{Obradors}},
  \bibinfo{author}{\bibfnamefont{W.~K.} \bibnamefont{Kwok}},
  \bibinfo{author}{\bibfnamefont{L.}~\bibnamefont{Paulius}},
  \bibinfo{author}{\bibfnamefont{G.~W.} \bibnamefont{Crabtree}},
  \bibnamefont{and}
  \bibinfo{author}{\bibfnamefont{G.}~\bibnamefont{Deutscher}},
  \bibinfo{journal}{Nat. Phys.} \textbf{\bibinfo{volume}{2}},
  \bibinfo{pages}{402} (\bibinfo{year}{2006}).

\bibitem[{\citenamefont{Moser et~al.}(1998)\citenamefont{Moser, Hug, Stiefel,
  and Guntherodt}}]{moser_1998}
\bibinfo{author}{\bibfnamefont{A.}~\bibnamefont{Moser}},
  \bibinfo{author}{\bibfnamefont{H.}~\bibnamefont{Hug}},
  \bibinfo{author}{\bibfnamefont{B.}~\bibnamefont{Stiefel}}, \bibnamefont{and}
  \bibinfo{author}{\bibfnamefont{H.}~\bibnamefont{Guntherodt}},
  \bibinfo{journal}{J. Magn. Magn. Matt.} \textbf{\bibinfo{volume}{190}},
  \bibinfo{pages}{114} (\bibinfo{year}{1998}).

\bibitem[{\citenamefont{Straver et~al.}(2008)\citenamefont{Straver, Hoffman,
  Auslaender, Rugar, and Moler}}]{straver_2008}
\bibinfo{author}{\bibfnamefont{E.~W.~J.} \bibnamefont{Straver}},
  \bibinfo{author}{\bibfnamefont{J.~E.} \bibnamefont{Hoffman}},
  \bibinfo{author}{\bibfnamefont{O.~M.} \bibnamefont{Auslaender}},
  \bibinfo{author}{\bibfnamefont{D.}~\bibnamefont{Rugar}}, \bibnamefont{and}
  \bibinfo{author}{\bibfnamefont{K.~A.} \bibnamefont{Moler}},
  \bibinfo{journal}{Appl. Phys. Lett.} \textbf{\bibinfo{volume}{93}},
  \bibinfo{pages}{172514} (\bibinfo{year}{2008}).

\bibitem[{\citenamefont{Auslaender et~al.}(2008)\citenamefont{Auslaender, Luan,
  Straver, Hoffman, Koshnick, Zeldov, Bonn, Liang, Hardy, and
  Moler}}]{Auslaender08}
\bibinfo{author}{\bibfnamefont{O.~M.} \bibnamefont{Auslaender}},
  \bibinfo{author}{\bibfnamefont{L.}~\bibnamefont{Luan}},
  \bibinfo{author}{\bibfnamefont{E.~W.~J.} \bibnamefont{Straver}},
  \bibinfo{author}{\bibfnamefont{J.~E.} \bibnamefont{Hoffman}},
  \bibinfo{author}{\bibfnamefont{N.~C.} \bibnamefont{Koshnick}},
  \bibinfo{author}{\bibfnamefont{E.}~\bibnamefont{Zeldov}},
  \bibinfo{author}{\bibfnamefont{D.~A.} \bibnamefont{Bonn}},
  \bibinfo{author}{\bibfnamefont{R.}~\bibnamefont{Liang}},
  \bibinfo{author}{\bibfnamefont{W.~N.} \bibnamefont{Hardy}}, \bibnamefont{and}
  \bibinfo{author}{\bibfnamefont{K.~A.} \bibnamefont{Moler}}
  (\bibinfo{year}{2008}), \bibinfo{note}{to be appear on Nat. Phys.}

\bibitem[{\citenamefont{Benkraouda and Clem}(1996)}]{Benkraouda1996}
\bibinfo{author}{\bibfnamefont{M.}~\bibnamefont{Benkraouda}} \bibnamefont{and}
  \bibinfo{author}{\bibfnamefont{J.~R.} \bibnamefont{Clem}},
  \bibinfo{journal}{Phys. Rev. B} \textbf{\bibinfo{volume}{53}},
  \bibinfo{pages}{438} (\bibinfo{year}{1996}).

\bibitem[{\citenamefont{Guikema et~al.}(2008)\citenamefont{Guikema, Bluhm,
  Bonn, Liang, Hardy, and Moler}}]{Guikema08}
\bibinfo{author}{\bibfnamefont{J.~W.} \bibnamefont{Guikema}},
  \bibinfo{author}{\bibfnamefont{H.}~\bibnamefont{Bluhm}},
  \bibinfo{author}{\bibfnamefont{D.~A.} \bibnamefont{Bonn}},
  \bibinfo{author}{\bibfnamefont{R.}~\bibnamefont{Liang}},
  \bibinfo{author}{\bibfnamefont{W.~N.} \bibnamefont{Hardy}}, \bibnamefont{and}
  \bibinfo{author}{\bibfnamefont{K.~A.} \bibnamefont{Moler}},
  \bibinfo{journal}{Phys. Rev. B} \textbf{\bibinfo{volume}{77}},
  \bibinfo{pages}{104515} (\bibinfo{year}{2008}).

\bibitem[{\citenamefont{Beleggia et~al.}(2004)\citenamefont{Beleggia, Pozzi,
  Tonomura, Kasai, Matsuda, Harada, Akashi, Masui, and Tajima}}]{Beleggia04}
\bibinfo{author}{\bibfnamefont{M.}~\bibnamefont{Beleggia}},
  \bibinfo{author}{\bibfnamefont{G.}~\bibnamefont{Pozzi}},
  \bibinfo{author}{\bibfnamefont{A.}~\bibnamefont{Tonomura}},
  \bibinfo{author}{\bibfnamefont{H.}~\bibnamefont{Kasai}},
  \bibinfo{author}{\bibfnamefont{T.}~\bibnamefont{Matsuda}},
  \bibinfo{author}{\bibfnamefont{K.}~\bibnamefont{Harada}},
  \bibinfo{author}{\bibfnamefont{T.}~\bibnamefont{Akashi}},
  \bibinfo{author}{\bibfnamefont{T.}~\bibnamefont{Masui}}, \bibnamefont{and}
  \bibinfo{author}{\bibfnamefont{S.}~\bibnamefont{Tajima}},
  \bibinfo{journal}{Phys. Rev. B} \textbf{\bibinfo{volume}{70}},
  \bibinfo{pages}{184518} (\bibinfo{year}{2004}).

\bibitem[{\citenamefont{Grigorenko et~al.}(2002)\citenamefont{Grigorenko,
  Bending, Koshelev, Clem, Tamegai, and Ooi}}]{Grigorenko2002}
\bibinfo{author}{\bibfnamefont{A.~N.} \bibnamefont{Grigorenko}},
  \bibinfo{author}{\bibfnamefont{S.~J.} \bibnamefont{Bending}},
  \bibinfo{author}{\bibfnamefont{A.~E.} \bibnamefont{Koshelev}},
  \bibinfo{author}{\bibfnamefont{J.~R.} \bibnamefont{Clem}},
  \bibinfo{author}{\bibfnamefont{T.}~\bibnamefont{Tamegai}}, \bibnamefont{and}
  \bibinfo{author}{\bibfnamefont{S.}~\bibnamefont{Ooi}},
  \bibinfo{journal}{Phys. Rev. Lett.} \textbf{\bibinfo{volume}{89}},
  \bibinfo{pages}{217003} (\bibinfo{year}{2002}).

\bibitem[{\citenamefont{Grigorenko et~al.}(2001)\citenamefont{Grigorenko,
  Bending, Tamegai, Ooi, and Henini}}]{Grigorenko2001}
\bibinfo{author}{\bibfnamefont{A.}~\bibnamefont{Grigorenko}},
  \bibinfo{author}{\bibfnamefont{S.}~\bibnamefont{Bending}},
  \bibinfo{author}{\bibfnamefont{T.}~\bibnamefont{Tamegai}},
  \bibinfo{author}{\bibfnamefont{S.}~\bibnamefont{Ooi}}, \bibnamefont{and}
  \bibinfo{author}{\bibfnamefont{M.}~\bibnamefont{Henini}},
  \bibinfo{journal}{Nature} \textbf{\bibinfo{volume}{414}},
  \bibinfo{pages}{728} (\bibinfo{year}{2001}).

\bibitem[{\citenamefont{Vlasko-Vlasov et~al.}(2002)\citenamefont{Vlasko-Vlasov,
  Koshelev, Welp, Crabtree, and Kadowaki}}]{vlasko-vlasov02}
\bibinfo{author}{\bibfnamefont{V.~K.} \bibnamefont{Vlasko-Vlasov}},
  \bibinfo{author}{\bibfnamefont{A.}~\bibnamefont{Koshelev}},
  \bibinfo{author}{\bibfnamefont{U.}~\bibnamefont{Welp}},
  \bibinfo{author}{\bibfnamefont{G.~W.} \bibnamefont{Crabtree}},
  \bibnamefont{and} \bibinfo{author}{\bibfnamefont{K.}~\bibnamefont{Kadowaki}},
  \bibinfo{journal}{Phys. Rev. B} \textbf{\bibinfo{volume}{66}},
  \bibinfo{pages}{014523} (\bibinfo{year}{2002}).

\bibitem[{\citenamefont{Liang et~al.}(1998)\citenamefont{Liang, Bonn, and
  Hardy}}]{Liang1998}
\bibinfo{author}{\bibfnamefont{R.}~\bibnamefont{Liang}},
  \bibinfo{author}{\bibfnamefont{D.}~\bibnamefont{Bonn}}, \bibnamefont{and}
  \bibinfo{author}{\bibfnamefont{W.}~\bibnamefont{Hardy}},
  \bibinfo{journal}{Physica C} \textbf{\bibinfo{volume}{304}},
  \bibinfo{pages}{105} (\bibinfo{year}{1998}).

\bibitem[{\citenamefont{Hosseini et~al.}(2004)\citenamefont{Hosseini, Broun,
  Sheehy, Davis, Franz, Hardy, Liang, and Bonn}}]{Hosseini04}
\bibinfo{author}{\bibfnamefont{A.}~\bibnamefont{Hosseini}},
  \bibinfo{author}{\bibfnamefont{D.~M.} \bibnamefont{Broun}},
  \bibinfo{author}{\bibfnamefont{D.~E.} \bibnamefont{Sheehy}},
  \bibinfo{author}{\bibfnamefont{T.~P.} \bibnamefont{Davis}},
  \bibinfo{author}{\bibfnamefont{M.}~\bibnamefont{Franz}},
  \bibinfo{author}{\bibfnamefont{W.~N.} \bibnamefont{Hardy}},
  \bibinfo{author}{\bibfnamefont{R.}~\bibnamefont{Liang}}, \bibnamefont{and}
  \bibinfo{author}{\bibfnamefont{D.~A.} \bibnamefont{Bonn}},
  \bibinfo{journal}{Phys. Rev. Lett.} \textbf{\bibinfo{volume}{93}},
  \bibinfo{pages}{107003} (\bibinfo{year}{2004}).

\bibitem[{\citenamefont{Broun et~al.}(2007)\citenamefont{Broun, Huttema,
  Turner, Ozcan, Morgan, Liang, Hardy, and Bonn}}]{Broun07}
\bibinfo{author}{\bibfnamefont{D.~M.} \bibnamefont{Broun}},
  \bibinfo{author}{\bibfnamefont{W.~A.} \bibnamefont{Huttema}},
  \bibinfo{author}{\bibfnamefont{P.~J.} \bibnamefont{Turner}},
  \bibinfo{author}{\bibfnamefont{S.}~\bibnamefont{Ozcan}},
  \bibinfo{author}{\bibfnamefont{B.}~\bibnamefont{Morgan}},
  \bibinfo{author}{\bibfnamefont{R.}~\bibnamefont{Liang}},
  \bibinfo{author}{\bibfnamefont{W.~N.} \bibnamefont{Hardy}}, \bibnamefont{and}
  \bibinfo{author}{\bibfnamefont{D.~A.} \bibnamefont{Bonn}},
  \bibinfo{journal}{Phys. Rev. Lett.} \textbf{\bibinfo{volume}{99}},
  \bibinfo{pages}{237003} (\bibinfo{year}{2007}).

\bibitem[{\citenamefont{Liang et~al.}(2005)\citenamefont{Liang, Bonn, Hardy,
  and Broun}}]{Liang05}
\bibinfo{author}{\bibfnamefont{R.}~\bibnamefont{Liang}},
  \bibinfo{author}{\bibfnamefont{D.~A.} \bibnamefont{Bonn}},
  \bibinfo{author}{\bibfnamefont{W.~N.} \bibnamefont{Hardy}}, \bibnamefont{and}
  \bibinfo{author}{\bibfnamefont{D.}~\bibnamefont{Broun}},
  \bibinfo{journal}{Phys. Rev. Lett.} \textbf{\bibinfo{volume}{94}},
  \bibinfo{pages}{117001} (\bibinfo{year}{2005}).

\bibitem[{\citenamefont{Albrecht et~al.}(1991)\citenamefont{Albrecht, Grutter,
  Horne, and Rugar}}]{Albrecht_1991}
\bibinfo{author}{\bibfnamefont{T.}~\bibnamefont{Albrecht}},
  \bibinfo{author}{\bibfnamefont{P.}~\bibnamefont{Grutter}},
  \bibinfo{author}{\bibfnamefont{D.}~\bibnamefont{Horne}}, \bibnamefont{and}
  \bibinfo{author}{\bibfnamefont{D.}~\bibnamefont{Rugar}}, \bibinfo{journal}{J.
  Appl. Phys.} \textbf{\bibinfo{volume}{69}}, \bibinfo{pages}{668}
  (\bibinfo{year}{1991}).

\bibitem[{Can()}]{Cantilever}
\bibinfo{note}{We used a commercial cantilever Nanosensors\texttrademark
  SSS-QMFMR with tip radius of curvature $\approx25$~nm.}

\bibitem[{Spr()}]{SpringConstant}
\bibinfo{note}{The spring constant of the cantilever is measured by Sader's
  method \cite{Sader1995}.}

\bibitem[{\citenamefont{Clem}(1994)}]{clem_1994}
\bibinfo{author}{\bibfnamefont{J.~R.} \bibnamefont{Clem}},
  \bibinfo{journal}{Physica C} \textbf{\bibinfo{volume}{235}},
  \bibinfo{pages}{2607} (\bibinfo{year}{1994}).

\bibitem[{dof()}]{doffset}
\bibinfo{note}{The micron-size $d_{\rm offset}$ mostly comes from the sample
  non-superconducting layer, which developed as a result of extensive surface
  cleaning due to contamination.}

\bibitem[{\citenamefont{Nelson and Vinokur}(1993)}]{Nelson1993}
\bibinfo{author}{\bibfnamefont{D.~R.} \bibnamefont{Nelson}} \bibnamefont{and}
  \bibinfo{author}{\bibfnamefont{V.~M.} \bibnamefont{Vinokur}},
  \bibinfo{journal}{Phys. Rev. B} \textbf{\bibinfo{volume}{48}},
  \bibinfo{pages}{13060} (\bibinfo{year}{1993}).

\bibitem[{\citenamefont{Nelson}(1988)}]{Nelson1988}
\bibinfo{author}{\bibfnamefont{D.~R.} \bibnamefont{Nelson}},
  \bibinfo{journal}{Phys. Rev. Lett.} \textbf{\bibinfo{volume}{60}},
  \bibinfo{pages}{1973} (\bibinfo{year}{1988}).

\bibitem[{\citenamefont{Reichhardt and Hastings}(2004)}]{olson_reichhardt2004}
\bibinfo{author}{\bibfnamefont{C.~J.~O.} \bibnamefont{Reichhardt}}
  \bibnamefont{and} \bibinfo{author}{\bibfnamefont{M.~B.}
  \bibnamefont{Hastings}}, \bibinfo{journal}{Phys. Rev. Lett.}
  \textbf{\bibinfo{volume}{92}}, \bibinfo{pages}{157002}
  (\bibinfo{year}{2004}).

\bibitem[{\citenamefont{Sader et~al.}(1999)\citenamefont{Sader, Chon, and
  Mulvaney}}]{Sader1995}
\bibinfo{author}{\bibfnamefont{J.}~\bibnamefont{Sader}},
  \bibinfo{author}{\bibfnamefont{J.}~\bibnamefont{Chon}}, \bibnamefont{and}
  \bibinfo{author}{\bibfnamefont{P.}~\bibnamefont{Mulvaney}},
  \bibinfo{journal}{Rev. Sci. Instrum.} \textbf{\bibinfo{volume}{70}},
  \bibinfo{pages}{3967} (\bibinfo{year}{1999}).

\end{thebibliography}
\end{document}